# Topolectrical-circuit realization of quadrupolar surface semimetals

Xingen Zheng[1], Tian Chen[*1], and Xiangdong Zhang[1+]

[1] Key Laboratory of advanced optoelectronic quantum architecture and measurements of Ministry of Education, Beijing Key Laboratory of Nanophotonics & Ultrafine Optoelectronic Systems, School of Physics, Beijing Institute of Technology, 100081, Beijing, China

[+*] Author to whom any correspondence should be addressed. E-mail: zhangxd@bit.edu.cn, chentian@bit.edu.cn

## Abstract

Recent studies have shown that there are two types of topological quadrupolar semimetals. One is the quadrupolar bulk semimetal, which is a conventional semimetal with gapless nodes in the bulk spectrum. The other type is named the quadrupolar surface semimetal, possessing a gapped bulk but gapless nodes on the surface spectrum. There have been many experiments on the implementation of quadrupolar bulk semimetals. However, the quadrupolar surface semimetal has never been realized. Here we report on the experimental realization of the quadrupolar surface semimetal in the circuit system. To highlight the properties of the quadrupolar surface semimetal, we also prepare a quadrupolar bulk semimetal in the circuit as a contrast. It has been demonstrated that the quadrupolar surface semimetal has a bulk band gap, but two gapless nodes appear on the surfaces. Also, the electrons not only concentrate on the hinges but also on the corresponding surfaces. Our studies provide novel ways to control electrical signals and may have potential applications in the field of intergraded circuit design.

## I. INTRODUCTION

In the past few decades, topological semimetal which refers to electronic gapless and exhibits topological band crossings around the Fermi level, has attracted great attention due to its series of novel properties [1-10]. Recently, higher-order topological semimetal with the higher-order topological state has drawn research interest again [11-17]. Unlike conventional first-order topological semimetals, three-dimensional (3D) higher-order topological semimetals possess one-dimensional hinge states. These topological semimetals have been fabricated experimentally in various systems such as electronic, photonic and acoustic systems [1-4, 18-20].

However, no matter the 1st-order or higher-order topological semimetal, what has been always concerned about is the semimetal with gapless nodes in the bulk. Here we call it bulk semimetal. Besides the traditional bulk semimetal, very recently another type of semimetal, which has a gap in the bulk but nodes on the surface, has been proposed theoretically [11]. Here it is called as 3D quadrupolar surface semimetal (QSSM). Such a type of semimetal has never been observed experimentally either.

In this work, we experimentally explore the QSSM by designing circuit networks. Recently, based on the similarity between circuit Laplacian and lattice Hamiltonian, simulating various physical phenomena with electric circuits has attracted lots of interest [21-33]. Compared with other classical platforms, circuit networks possess remarkable advantages of being versatile and reconfigurable. So far, many extremely complex topological states have been observed using circuit networks [34-44]. In the following, we discuss the experimentally feasible scheme for realizing the 3D QSSM based on the electric circuits. Meanwhile, we also construct a quadrupolar bulk semimetal (QBSM) as a contrast.

## II. THEORETICAL DESIGN OF TOPOLECTRICAL-CIRCUIT QUADRUPOLAR SEMIMETALS

On the one hand, the quadrupolar topological semimetals virtually can be understood and designed by stacking the quadrupolar topological insulators in layers with proper vertical couplings [11]. This design theory is shown in appendix B. On the other hand, in the theory of simulating tight-binding model by the circuit, the positive and negative tunneling strength between lattice sites are replaced by inductor and capacitor, and every site is given proper grounding elements to offset the degree matrix. Then, at the resonate frequency $\omega_0$, the admittance matrix describing the circuit is exactly equivalent to the Hamiltonian describing the tight-binding model, with only one coefficient difference [24].

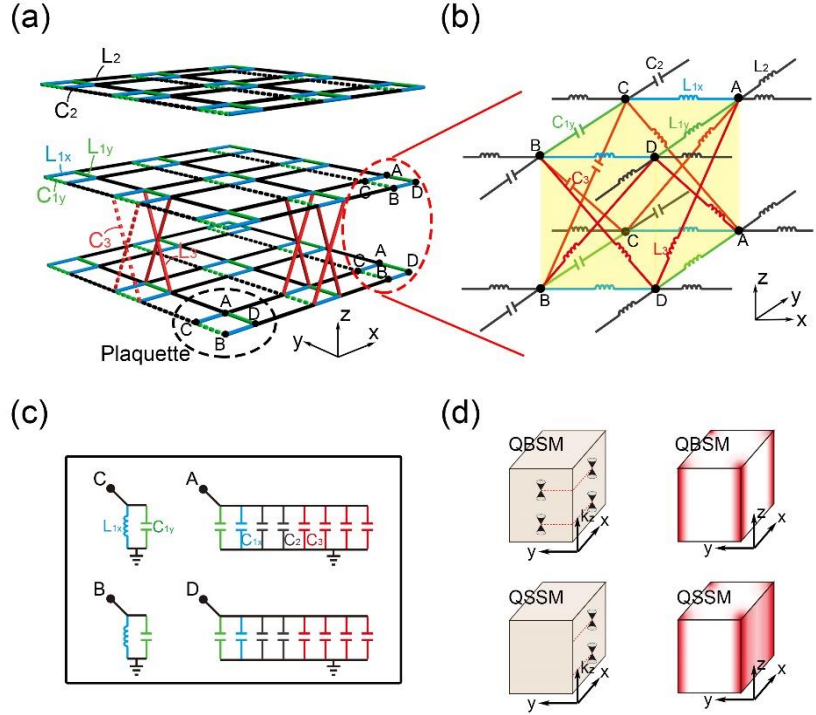

FIG. 1. (a) Circuit scheme with $3\times3\times3$ plaquettes. (b) Detailed circuit connections of two plaquettes in adjacent layers. (c) The grounding elements for one plaquette under the fully periodic boundary condition. (d) Left panels are the nodes that can be found on the surfaces of the two semimetals; Right panels are the electron density of two semimetals.

Based on these theories, we fabricate a circuit to implement QBSM and QSSM, as shown in Fig. 1(a). This circuit structure corresponds to a lattice model. The plaquette in the figure corresponds to the unit cell in the lattice. And we choose the lattice constant $a_{x,y,z}=1$, so the basic vectors in three directions are (1,0,0), (0,1,0) and (0,0,1). The solid and dashed cylinders with different colors represent different inductors and capacitors that are marked in the figure. Here, although we only show one circuit, both QBSM and QSSM can be represented by it for these two semimetals have similar coupling patterns. The difference between the two circuit semimetals lies in the values of the circuit elements. Every plaquette in the circuit has four sites

A, B, C and D, which are depicted in the black dashed circle in Fig. 1(a). The coupling relationship between sites represented by capacitors and inductors is shown in the figure. Notice that the vertical coupling terms ($C_3$ and $L_3$ in the circuit) exist in every plaquette where some of them are omitted to avoid clutter. To show the coupling relationship clearly, detailed circuit connections of two plaquettes in adjacent layers are given in Fig. 1(b), corresponding to the red dashed circle in Fig. 1(a). Under the fully periodic boundary condition, the grounding elements for every plaquette are shown in Fig. 1(c) which are omitted in Figs. 1(a) and 1(b). But when the open boundary condition is chosen, the grounding elements on the edge should be adjusted properly to keep the diagonal matrix elements of the circuit Laplacian unchanged. Even though this circuit is only a schematic diagram with $3 \times 3 \times 3$ plaquettes, it's enough to show the arrangements of plaquettes at the edges and bulk of the lattice. And any circuit of $N_x \times N_y \times N_z$ plaquettes can be obtained by a simple extension.

The electronic elements in the circuit diagram as shown in Fig. 1(a) are satisfied as, $C_{1x} : C_{1y} : C_2 : C_3 = \gamma_x : \gamma_y : \lambda : \chi$ and $1/L_{1x} : 1/L_{1y} : 1/L_2 : 1/L_3 = \gamma_x : \gamma_y : \lambda : \chi$ (here we stipulate $\lambda = 1$). Using Kirchhoff's current law, the circuit Laplacian $J(\mathbf{k})$ for both semimetals in the moment space at the resonance frequency $\omega_0 = 1/\sqrt{L_2 C_2}$ can be derived as

$$J(\mathbf{k}) = i\sqrt{\frac{C_2}{L_2}}[(\gamma_x + 2\chi \cos k_z + \lambda \cos k_x)\Gamma_4 + \lambda \sin k_x \Gamma_3 \qquad (1)$$
$$+ (\gamma_y + 2\chi \cos k_z + \lambda \cos k_y)\Gamma_2 + \lambda \sin k_y \Gamma_1],$$

where $\Gamma_0 = \tau_3 \otimes I_2$, $\Gamma_k = -\tau_2 \otimes \sigma_k$ and $\Gamma_4 = \tau_1 \otimes I_2$, $I_2$ is the $2 \times 2$ identity matrix, and $\tau_\alpha$ and $\sigma_k$ are Pauli matrices. $k_i$ (i=x,y,z) is the phase of Block wave vector propagating along the x, y, and z directions, respectively. The derivation process of the admittance matrix is given in the Appendix A. Then, the corresponding Hamiltonian of the tight-binding model can be regarded as $H(\mathbf{k}) = J(\mathbf{k})/(i\sqrt{C_2/L_2})$, which is exactly the lattice counterpart of this circuit structure. In the experiment, to implement the QBSM, we take $\gamma_i = \lambda = 1$ (i=x,y) and $\chi = 1/3.3$. To implement the QSSM, we take $\gamma_x = \lambda = 1$, $\gamma_y = 1/3.3$ and $\chi = 1/3.3$. The only difference between the two semimetals is the value of $\gamma_y$ (the reason why we choose these parameters is shown in Appendix B). It is seen that, for both semimetals, x and y mirror symmetries are satisfied with representation matrices $\hat{m}_x = \tau_1 \otimes \sigma_3$ and $\hat{m}_y = \tau_1 \otimes \sigma_1$, which obey the anticommute relationship $\{\hat{m}_x, \hat{m}_y\} = 0$. Moreover, the QBSM has the $C_4$ rotation symmetry with the representation

$$r_4 = \begin{pmatrix} 0 & I_2 \\ -i\sigma_2 & 0 \end{pmatrix}, \qquad (2)$$

but the QSSM breaks this $C_4$ rotation symmetry by $\gamma_x \neq \gamma_y$.

Before executing the experiments for the two semimetals, some electric simulations with the parameters of the two semimetals described above are put forward. For the QBSM, two gapless nodes in the bulk spectrum appear as well as in the surfaces normal to x and y (as shown in the left panel of Fig. 1(d)). While for the QSSM, it has a bulk band gap, but two gapless nodes appear on the surface spectrum normal to $\hat{y}$, which is novel and has not been observed in the experiment before. Besides the nodes in the spectrum, the electron density is also different from the traditional QBSM. Here, we choose the periodic boundary condition along z direction, and open boundary conditions for x and y directions. For the QBSM, the electrons mainly concentrate on the hinges because of the nonzero quadrupole moment as shown in the right panel of Fig. 1(d). However, for the QSSM, the electrons not only concentrate on the hinges but also the surfaces normal to y. To clearly show the location of the nodes, we also give the simulated admittance spectrums by solving the circuit Laplacian J(k) of the two semimetals. For the QBSM, we take $C_{1x}=C_{1y}=C_2=3.3nF$, $C_3=1nF$, $L_{1x}=L_{1y}=L_2=100uH$ and $L_3=330uH$. We find two nodes at the location ($k_x=\pi$, $k_y=\pi$, $k_z=\pi/2$) and ($k_x=\pi$, $k_y=\pi$, $k_z=-\pi/2$) in the bulk spectrum as shown in the 2D subspace formed by $k_x$ and $k_y$ with $k_z=\pi/2$ in Fig. 2(a). These two nodes can be also found when projected to the surface normal to y, as shown in Fig. 2(b). For the QSSM, we take $C_{1x}=C_2=3.3nF$, $C_{1y}=C_3=1nF$, $L_{1x}=L_2=100uH$ and $L_{1y}=L_3=330uH$. Then we get a gapped spectrum in the bulk as shown in Fig. 2(c), but two nodes locating at ($k_x=\pi$, $k_z=\pi/2$) and ($k_x=\pi$, $k_z=-\pi/2$) in the surface normal to y as shown in Fig. 2(d). Thus, the QSSM and QBSM phases should be observed experimentally by designing the quadrupolar semimetal circuit.

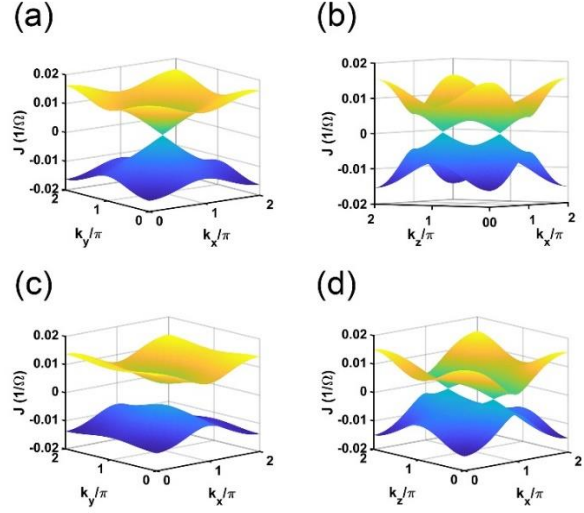

**FIG. 2. (a) Spectrum of the bulk of quadrupolar bulk semimetal. (b) Spectrum of the surface normal to y of quadrupolar bulk semimetal. (c) Spectrum of the bulk of quadrupolar surface semimetal. (d) Spectrum of the surface normal to y of quadrupolar surface semimetal.**

## III. EXPERIMENTAL OBSERVATIONS OF QUADRUPOLAR BULK AND SURFACE SEMIMETALS IN THE CIUCUIT SYSTEM

In the experiments, we prepare a circuit with $4\times 4\times 4$ plaquettes by simply extending the circuit scheme in Fig. 1(a). This is because we want to measure the admittance spectrum (corresponds to the energy spectrum of the lattice model) of the circuit in the momentum space and observe the gapless bulk or surface nodes on the band structure. According to the method of spectrum-measuring in the circuit, the resolution of momentum space is determined by the size of the circuit [45]. In order to observe the key points of $k_i=\pi/2$ in the momentum space, the circuit should have $4\times 4\times 4$ plaquettes at least.

The whole experimental sample is shown in Fig. 3(a). It is composed of four layers of Printed Circuit Boards (PCBs). Each layer contains $4\times 4$ plaquettes. The copper pillar in the experimental sample is used to support four layers of PCBs, which is also connected to the ground of the circuit. The couplings of electric elements between two adjacent layers are arranged on the lower layer, and the pin header connectors are prepared on every PCB so that two adjacent layers could connect with each other through the cables. To see more clearly, a detailed front look and back look of one plaquette of the circuit is given in Figs. 3(b) and 3(c), corresponding to the red dashed rectangle in Fig. 3(a). In Fig. 3(b), A, B, C and D are four sites within a plaquette, while A' B' C' and D' represent the sites that should be connected to the next layer. Through pin header connectors on each site, when the sites A', B', C' and D' in this layer are connected to the sites A, B, C and D in the next layer, then two adjacent layers are coupled. The line-via hole is used to pass the wire connecting the top and bottom layers, because the periodic boundary condition along the z direction is always applied in our experiment. Whereas the periodic and open boundary conditions along x and y directions should be switched according to the observation requirement, which is achieved by adjusting the electronic elements and groundings on the edge. The grounding elements of the plaquette can be seen in Fig. 3(c), which is on the back of the PCB.

In the circuit, there exist various stray capacitance and inductance, in which the maximum error is the parasitic inductance of the long wires. Therefore, to resist the parasitic inductance, the value of inductors we use is large enough (more than 100uH). And all the circuit elements have a tolerance of 5%.

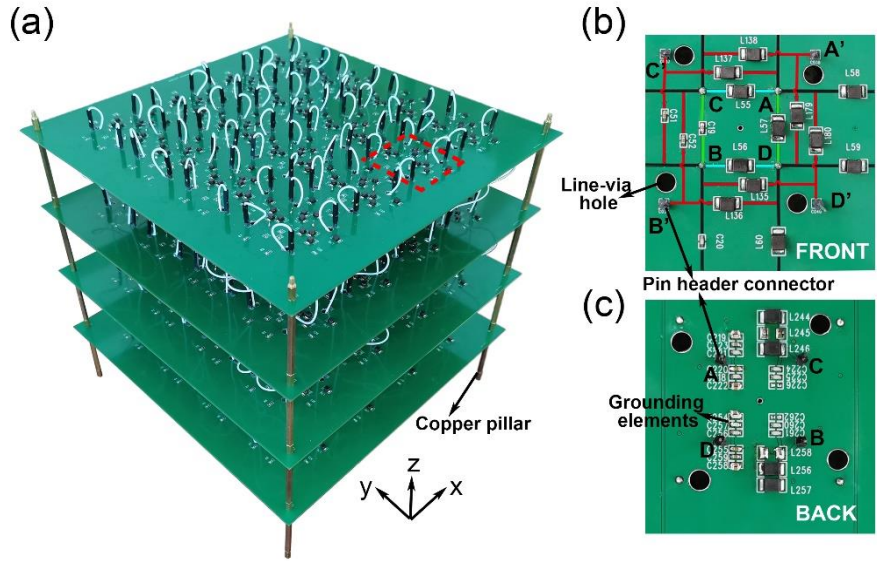

FIG. 3. (a) The whole view of the experimental setup. (b) The front view of a plaquette. (c) The back view of the plaquette.

To prove these two circuit networks really realize the QBSM and QSSM, we measure the admittance spectrum of the circuits in the momentum space to see whether there are nodes on the bulk and surface spectrum. The admittance spectrum can be obtained experimentally by an elementary voltage measurement and the Fourier transform [45]. Detailed spectrum-measurement procedures and the experimental instruments we use see appendix C. For the QBSM, after the periodic boundary conditions are applied in x, y and z directions, two nodes at the location ($k_x=\pi$, $k_y=\pi$, $k_z=\pi/2$) and ($k_x=\pi$, $k_y=\pi$, $k_z=-\pi/2$) can be observed. We show the spectrum measured in the 2D subspace formed by $k_x$ and $k_y$ with $k_z=\pi/2$ in Fig. 4(a), where the red dots are measured results and the translucent surfaces are theoretical simulations. For the QSSM, the periodic boundary conditions are only applied in the x and z directions. The measured spectrum is shown in Fig. 4(d) where two nodes at the location ($k_x=\pi$, $k_z=\pi/2$) and ($k_x=\pi$, $k_z=-\pi/2$) are observed. The experimental results agree well with the simulations in Fig. 2.

To prove they are quadrupolar semimetals, we choose the open boundary conditions for the x and y directions in the circuits but keep the periodic boundary condition for the z direction. Then, The spectrums of the circuit Laplacian $J(\omega)$ as a function of the driving frequency $\omega$ are theoretically obtained, as shown in Figs. 4(b) and 4(e). We find that for both QBSM and QSSM, there exists a zero-energy mode near the resonate frequency $\omega=\omega_0$ as shown in Figs. 4(b) and 4(e), which corresponds to the hinge state of the quadrupolar semimetal. Besides, the impedance in the QBSM and QSSM circuits are provided in Figs. 4(c) and 4(f). The impedance $Z_{ab}$ between two nodes a and b can be expressed as

$$Z_{ab}(\omega) = \sum_n \frac{|\phi_n(a)-\phi_n(b)|^2}{j_n(\omega)}, \tag{4}$$

where $j_n(\omega)$ is the nth eigenvalue of the circuit Laplacian J, and $\phi_n(i)$ is the component of the nth eigenstate at node i (i=a,b). Usually, node b in Eq. (4) is chosen to be the ground. In this case, for the zero-energy edge state, the zero eigenvalue corresponds to a very small denominator in Eq. (4), and the strong localization of the eigenstate at the edge brings about a very large numerator in Eq. (4). Then, a large impedance appears. Therefore, according to this relation, the relative impedance at each node can approximately be used to reflect the relative electron density of the zero-energy state in the lattice model, and the hinge state can be directly observed which is the characteristic of the quadrupolar semimetals. We show the impedance of the nodes at the hinge, the surface normal to x/y and the bulk in Figs. 4(c) and 4(f) for the QBSM and QSSM, respectively. The coordinates of which nodes we measure specifically are provided in appendix C. For both circuits, we find that near the resonate frequency $\omega_0=277 kHz$, the impedance at the hinge is very large, and that at the bulk is small. Besides the similarities of hinge modes for both two semimetals, there are also differences between QBSM and QSSM on the surfaces. For the QBSM, both surfaces normal to x and y have a moderate impedance. But for the QSSM, the surface normal to y has a large impedance compared to the surface normal to x. This impedance distribution, which is also corresponded to the electron density distribution shown in the insets of Figs. 4(c) and 4(f), illustrates the hinge state in both semimetals and the surface concentration in the QSSM.

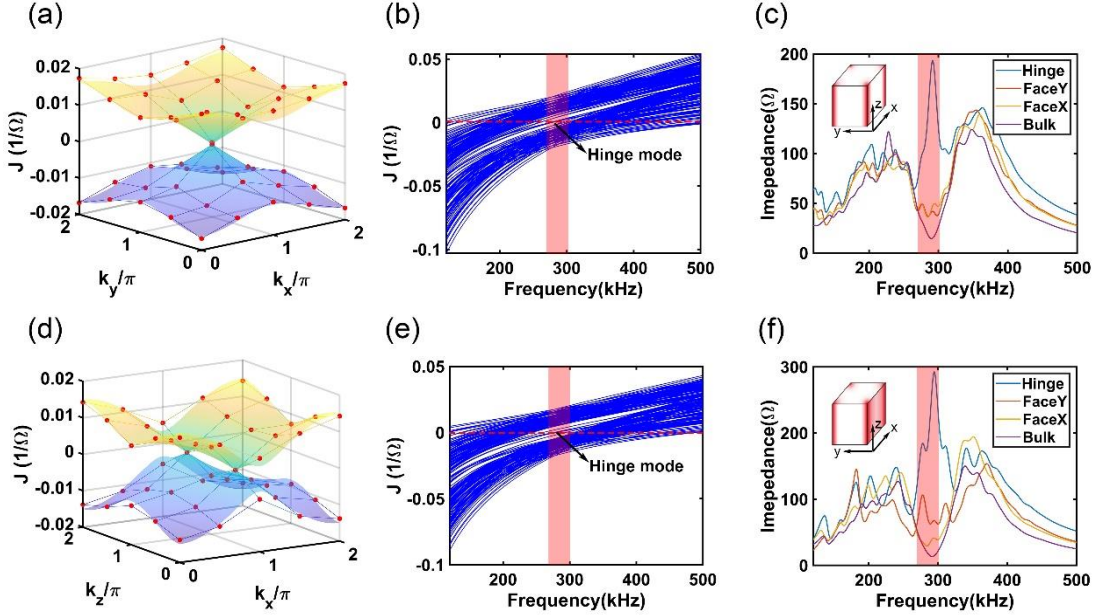

**FIG. 4. (a)-(c) and (d)-(e) are corresponded to the quadrupolar bulk and surface semimetals, respectively. (a) and (d) are the measured admittance spectrums in the momentum space. (b) and (e) are the spectrums of the circuit Laplacian as a function of the driving frequency. (c) and (f) are the impedances measured at the hinge, the surface and the bulk. The surfaces normal to x and y are marked as FaceX and FaceY.**

## IV. CONCLUSION

In conclusion, we have constructed 3D circuit networks to realize a classical analog of the QBSM. For comparison, the topolectrical-circuit realization of the QBSM has also been provided. Our circuit implementations have confirmed the characteristics of QSSM, including gapless nodes on the surface spectrum, the hinge states and the electrons concentration on the surface. We believe that our research has potential applications in a variety of disciplines. For example, our circuit construction method may be applied to other models; the layer-stacking method in our work may be used to realize other types of topological semimetals. Our study paves the way for employing topolectrical-circuit to study complex phenomena of topological semimetals and offers possibilities to control electrical signals in unprecedented ways.

## ACKNOWLEDGMENTS

This work is supported by the National Key R & D Program of China under Grant No. 2017YFA0303800 and the National Natural Science Foundation of China (No. 91850205, and No. 12104041).

## APPENDIX A: DERIVATION OF CIRCUIT LAPLACIAN $J$ IN THE MOMENTUM SPACE

The voltages on sites A, B, C and D shown in Figs. 1(b)-1(d) are labeled by $V_A$, $V_B$, $V_C$ and $V_D$, respectively. The corresponding input currents are described by $I_A$, $I_B$, $I_C$ and $I_D$. The phases of the wave vector propagating along the $x$, $y$ and $z$ directions are denoted by $k_x$, $k_y$, and $k_z$, respectively. After applying Kirchhoff's current law to the four nodes, we have

$$\begin{aligned}
I_A &= \frac{1}{i\omega L_2}(V_A - V_C e^{ik_x}) + \frac{1}{i\omega L_{1x}}(V_A - V_C) + \frac{1}{i\omega L_2}(V_A - V_D e^{ik_y}) + \frac{1}{i\omega L_{1y}}(V_A - V_D) + \frac{1}{i\omega L_3}(V_A - V_D e^{ik_z}) + \frac{1}{i\omega L_3}(V_A - V_C e^{ik_z}) \\
&\quad + \frac{1}{i\omega L_3}(V_A - V_D e^{-ik_z}) + \frac{1}{i\omega L_3}(V_A - V_C e^{-ik_z}) + i\omega C_{1x}(V_A - 0) + i\omega C_{1y}(V_A - 0) + 2i\omega C_2(V_A - 0) + 4i\omega C_3(V_A - 0) \\
I_B &= \frac{1}{i\omega L_{1x}}(V_B - V_D) + \frac{1}{i\omega L_2}(V_B - V_D e^{-ik_x}) + i\omega C_{1y}(V_B - V_C) + i\omega C_2(V_B - V_C e^{-ik_y}) + i\omega C_3(V_B - V_C e^{ik_z}) + \frac{1}{i\omega L_3}(V_B - V_D e^{ik_z}) \\
&\quad + i\omega C_3(V_B - V_C e^{-ik_z}) + \frac{1}{i\omega L_3}(V_B - V_D e^{-ik_z}) + \frac{1}{i\omega L_{1x}}(V_B - 0) + i\omega C_{1y}(V_B - 0) \\
I_C &= \frac{1}{i\omega L_{1x}}(V_C - V_A) + \frac{1}{i\omega L_2}(V_C - V_A e^{-ik_x}) + i\omega C_{1y}(V_C - V_B) + i\omega C_2(V_C - V_B e^{ik_y}) + i\omega C_3(V_C - V_B e^{ik_z}) + \frac{1}{i\omega L_3}(V_C - V_A e^{ik_z}) \\
&\quad + i\omega C_3(V_C - V_B e^{-ik_z}) + \frac{1}{i\omega L_3}(V_C - V_A e^{-ik_z}) + \frac{1}{i\omega L_{1x}}(V_B - 0) + i\omega C_{1y}(V_B - 0) \\
I_D &= \frac{1}{i\omega L_2}(V_D - V_B e^{ik_x}) + \frac{1}{i\omega L_{1x}}(V_D - V_B) + \frac{1}{i\omega L_2}(V_D - V_A e^{-ik_y}) + \frac{1}{i\omega L_{1y}}(V_D - V_A) + \frac{1}{i\omega L_3}(V_D - V_A e^{ik_z}) + \frac{1}{i\omega L_3}(V_D - V_B e^{ik_z}) \\
&\quad + \frac{1}{i\omega L_3}(V_D - V_A e^{-ik_z}) + \frac{1}{i\omega L_3}(V_D - V_B e^{-ik_z}) + i\omega C_{1x}(V_A - 0) + i\omega C_{1y}(V_A - 0) + 2i\omega C_2(V_A - 0) + 4i\omega C_3(V_A - 0)
\end{aligned}$$

(A1)

Eq. (A1) can be written in the form of $\mathbf{I} = J\mathbf{V}$. We set the driving frequency as $\omega = \omega_0 = 1/\sqrt{L_2 C_2}$, with $C_{1x} : C_{1y} : C_2 : C_3 = \gamma_x : \gamma_y : \lambda : \chi$ and $1/L_{1x} : 1/L_{1y} : 1/L_2 : 1/L_3 = \gamma_x : \gamma_y : \lambda : \chi$ (here we stipulate $\lambda = 1$). On this resonance frequency $\omega_0 = 1/\sqrt{L_2 C_2}$ (also equals to $1/\sqrt{L_{1x} C_{1x}} = 1/\sqrt{L_{1y} C_{1y}} = 1/\sqrt{L_3 C_3}$), we have $i\omega C_{1x} = -1/i\omega L_{1x}$, $i\omega C_{1y} = -1/i\omega L_{1y}$, $i\omega C_2 = -1/i\omega L_2$ and $i\omega C_3 = -1/i\omega L_3$. Substituting these relations into Eq. (A1), the diagonal elements of the circuit Laplacian $J$ all cancel out, and $J$ could be written as

$$J = i\sqrt{\frac{C_2}{L_2}} \begin{pmatrix} 0 & 0 & \gamma_x + 2\chi \cos k_z + \lambda e^{ik_x} & \gamma_y + 2\chi \cos k_z + \lambda e^{ik_y} \\ 0 & 0 & -\gamma_y - 2\chi \cos k_z - \lambda e^{-ik_y} & \gamma_x + 2\chi \cos k_z + \lambda e^{-ik_x} \\ \gamma_x + 2\chi \cos k_z + \lambda e^{-ik_x} & -\gamma_y - 2\chi \cos k_z - \lambda e^{ik_y} & 0 & 0 \\ \gamma_y + 2\chi \cos k_z + \lambda e^{-ik_y} & \gamma_x + 2\chi \cos k_z + \lambda e^{ik_x} & 0 & 0 \end{pmatrix}.$$

(A2)

When we set $\Gamma_0 = \tau_3 \otimes I_2$, $\Gamma_k = -\tau_2 \otimes \sigma_k$ and $\Gamma_4 = \tau_1 \otimes I_2$, where $I_2$ is the $2 \times 2$ identity matrix, and $\tau_\alpha$ and $\sigma_k$ are Pauli matrices. Eq. (A2) can be transformed to Eq. (1) in the main text.

### APPENDIX B: THE DESIGN THEORY OF THE QUADRUPOLAR BULK AND SURFACE SEMIMETALS

First, let us review the model of a 3D topological quadrupolar semimetal [11]. With four spinless orbitals per unit cell, the tight-binding Hamiltonian of the 3D topological quadrupolar semimetal can be expressed in Bloch representation:

$$H(\mathbf{k}) = [\gamma_x + \chi_x(k_z) + \lambda \cos k_x]\Gamma_4 + \lambda \sin k_x \Gamma_3 \\ + [\gamma_y + \chi_y(k_z) + \lambda \cos k_y]\Gamma_2 + \lambda \sin k_y \Gamma_1, \quad \text{(B1)}$$

where $\Gamma_0 = \tau_3 \otimes I_2$, $\Gamma_k = -\tau_2 \otimes \sigma_k$ and $\Gamma_4 = \tau_1 \otimes I_2$, $I_2$ is the $2 \times 2$ identity matrix, and $\tau_\alpha$ and $\sigma_k$ are Pauli matrices. If $\chi_i = 0$, Eq. (B1) corresponds to the Hamiltonian of the 2D quadrupolar insulator with $\gamma_i$ and $\lambda_i$ the intra- and inter-cell nearest-neighbor tunneling strength. This 2D quadrupolar insulator exhibits a $Z_2 \times Z_2$ set of topological classes whose phase diagram is shown in Fig. B1. The location of the point ($\gamma_x/|\lambda|$, $\gamma_y/|\lambda|$) represents the topological classes of the 2D quadrupolar insulator. Any ($\gamma_x/|\lambda|$, $\gamma_y/|\lambda|$) has $x$ and $y$ mirror symmetries with representation matrices $\hat{m}_x = \tau_1 \otimes \sigma_3$ and $\hat{m}_y = \tau_1 \otimes \sigma_1$, but has $C_4$ rotation symmetry only on the diagonal line of the phase diagram with the representation as shown in Eq. (2) of the main text. There is $\pi$ flux per plaquette, so the two mirror symmetric operators anticommute with each other and $r_4^4 = -1$. Based on this 2D quadrupolar insulator, the 3D TQSM can be constructed when properly defining the $\chi_i(k_z)$ as periodic functions on the $k_z$ Brillouin zone (BZ) by stacking the 2D quadrupoles and coupling the layers. In this case, the quantities $\gamma_i(k_z) \equiv \gamma_i + \chi_i(k_z)$ in Eq. (1) represent maps from the $k_z$ BZ to closed paths in the 2D phase diagram Fig. B1. When the paths go through the transition point between two different topological classes, gapless bulk or surface nodes appear in the momentum space, which is exactly the symbol of a semimetal.

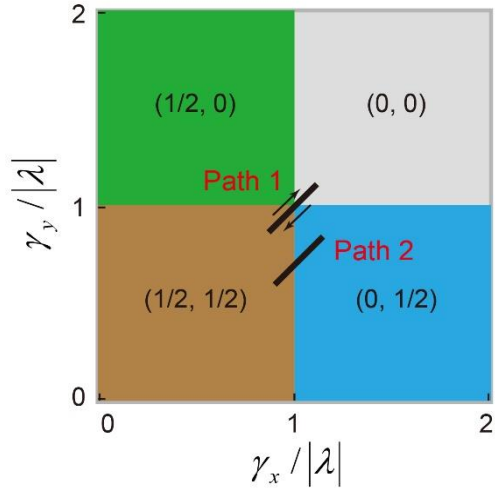

**FIG. B1. The phase diagram of the 2D quadrupolar insulator as a function of ($\gamma_x/|\lambda|$, $\gamma_y/|\lambda|$). The two paths correspond to the QBSM and QSSM in our experiments.**

The two quadrupolar semimetals in our experiment correspond to Path 1 and Path 2 in the phase diagram as shown in Fig. B1. For the path 1, $\gamma_i = \lambda = 1$ and $\chi_i = 2\cos k_z / 3.3$. This path corresponds to the QBSM in our main text. When it passes through the $C_4$ invariant phase transition point when $k_z = \pm\pi/2$. Two bulk gapless nodes in the bulk energy spectrum appear. For the path 2, $\gamma_x = \lambda = 1$, $\gamma_y = 1/3.3$ and $\chi_i = 2\cos k_z/3.3$. This path passes through the phase transition point between the topological class (1/2, 1/2) and (0, 1/2) when $k_z = \pm\pi/2$. Therefore, this path corresponds to the QSSM which has a bulk band gap, but two gapless nodes on the surfaces normal to $\hat{y}$ when $k_z = \pm\pi/2$.

## APPENDIX C: DETAILS FOR THE SPECTRUM-MEASUREMENT AND OTHER EXPERIMENT THINGS

First, we introduce how to reconstruct the admittance spectrum in the momentum space by an elementary voltage measurement and Fourier transform. For a circuit with $N$ nodes, the current flowing into the circuit at node $i$ is denoted by $I_i$, and the voltage at node $j$ is denoted by $V_j$. After arranging the components $I_i$ and $V_j$ in a vector form, we get $\mathbf{I} = J(\mathbf{r})\mathbf{V}$ following Kirchhoff's law. $J(\mathbf{r})$ is the circuit Laplacian in the real space. The method of reconstructing the admittance spectrum in the momentum space is actually measuring the circuit Laplacian $J(\mathbf{r})$ of the circuit in real space, then through the Fourier transform, the circuit Laplacian $J(\mathbf{k})$ in the momentum space is obtained, and we can calculate the admittance spectrum. So now the goal is how to obtain the circuit Laplacian $J(\mathbf{r})$ by measuring the circuit.

We left-multiply both sides of this equation $\mathbf{I} = J(\mathbf{r})\mathbf{V}$ by $G$ which is the inverse matrix of $J$, and we get $G\mathbf{I} = \mathbf{V}$. Here $G$ is called the impedance matrix. Then we excite only one node of the circuit such as at node $j$, and then measure the input current $I_j$ and the voltages $V_i$ of all $N$ nodes. One column of the impedance matrix can be obtained by $G_{ij} = V_i/I_j$. We excite the remaining nodes in the circuit and repeat the measurement of the input current and voltages for $N$ nodes. After all these operations, we get the whole impedance matrix $G$. Finally, we take the inverse of $G$ and get $J$. Then through Fourier transformation, $J(\mathbf{k})$ could be obtained. Note that usually there exist symmetries in the model and we needn't to excite all nodes. For example, if the model is all periodic, it has the translational symmetry. And we just need to excite four nodes A, B, C, D in one certain plaquette, because all plaquettes are translationally symmetric.

Second, we explain which nodes are measured specifically when we demonstrate the impedance at the bulk, surface and hinge in the main text. Here, we use (x,y,z,t) to denote the specific node, where x, y and z represent the coordination of a plaquette and t (t=A, B, C and D) represents which node of this plaquette. For quadrupolar bulk semimetal, the measured nodes corresponding to hinge, surface normal to x, surface normal to y and bulk are (4,4,4,A), (4,3,4,A), (3,1,4,D) and (3,3,4,A) respectively. For quadrupolar surface semimetal, the measured nodes corresponding to hinge, surface normal to x, surface normal to y, and bulk are (4,4,4,A), (4,3,4,A) and (2,4,4,C) and (3,3,4,A) respectively. If we don't measure these nodes but other nodes at the hinge, surface and bulk, the results are similar.

Third, we introduce the instrument used in the experiment. We use the Wayne Kerr 6500B precision impedance analyzer to measure the impedance of the circuit as a function of the driving frequency. When measuring the spectrum of the circuit, we use the RIGOL DG1022U signal generator as the power source, and Agilent DSO7104B oscilloscope to measure the voltage at each node.

**Reference**

[1] N. P. Armitage, E. J. Mele, and A. Vishwanath, Weyl and Dirac semimetals in three-dimensional solids, Rev. Mod. Phys. **90**, 015001 (2018).

[2] S.-Y. Xu *et al.*, Discovery of a Weyl fermion semimetal and topological Fermi arcs, Science **349**, 613 (2015).

[3] A. Bernevig, H. Weng, Z. Fang, and X. Dai, Recent progress in the study of topological semimetals, J. Phys. Soc. Jpn. **87**, 041001 (2018).

[4] A. A. Burkov, Topological semimetals, Nat. Mater. 15, 1145 (2016).

[5] X. Wan, A. M. Turner, A. Vishwanath, and S. Y. Savrasov, Topological semimetal and Fermi-arc surface states in the electronic structure of pyrochlore iridates, Phys. Rev. B **83**, 205101 (2011).

[6] H. Weng, C. Fang, Z. Fang, B. A. Bernevig, and X. Dai, Weyl Semimetal Phase in Noncentrosymmetric Transition-Metal Monophosphides, Phys. Rev. X **5**, 011029 (2015).

[7] S. Ma, Y. Bi, Q. Guo, B. Yang, O. You, J. Feng, H.-B. Sun, and S. Zhang, Linked Weyl

surfaces and Weyl arcs in photonic metamaterials, Science 373, 572 (2021).

[8] A. A. Soluyanov, D. Gresch, Z. Wang, Q. Wu, M. Troyer, X. Dai, and B. A. Bernevig, Type-II Weyl semimetals, Nature 527, 495 (2015).

[9] J. Hu, S.-Y. Xu, N. Ni, and Z. Mao, Transport of topological semimetals, Ann. Rev. Mater. Res. 49, 207 (2019).

[10] B. Q. Lv, T. Qian, and H. Ding, Experimental perspective on three-dimensional topological semimetals, Rev. Mod. Phys. 93, 025002 (2021).

[11] M. Lin and T. L. Hughes, Topological quadrupolar semimetals, Phys. Rev. B **98**, 241103 (2018).

[12] H.-X. Wang, Z.-K. Lin, B. Jiang, G.-Y. Guo, and J.-H. Jiang, Higher-Order Weyl Semimetals, Phys. Rev. Lett. **125**, 146401 (2020).

[13] S. A. A. Ghorashi, T. Li, and T. L. Hughes, Higher-Order Weyl Semimetals, Phys. Rev. Lett. **125**, 266804 (2020).

[14] M. Ezawa, Higher-Order Topological Insulators and Semimetals on the Breathing Kagome and Pyrochlore Lattices, Phys. Rev. Lett. **120**, 026801 (2018).

[15] R.-X. Zhang, Y.-T. Hsu, and S. Das Sarma, Higher-order topological Dirac superconductors, Phys. Rev. B **102**, 094503 (2020).

[16] Q.-B. Zeng, Y.-B. Yang, and Y. Xu, Higher-order topological insulators and semimetals in generalized Aubry-Andre-Harper models, Phys. Rev. B **101**, 241104 (2020).

[17] T. Liu, J. J. He, Z. Yang, and F. Nori, Higher-Order Weyl-Exceptional-Ring Semimetals, Phys. Rev. Lett. **127**, 196801 (2021).

[18] Q. Wei, X. Zhang, W. Deng, J. Lu, X. Huang, M. Yan, G. Chen, Z. Liu, and S. Jia, Higher-order topological semimetal in acoustic crystals, Nat. Mater. **20**, 812 (2021).

[19] L. Luo, H.-X. Wang, Z.-K. Lin, B. Jiang, Y. Wu, F. Li, and J.-H. Jiang, Observation of a phononic higher-order Weyl semimetal, Nat. Mater. **20**, 794 (2021).

[20] Z. Wang, D. Liu, H. T. Teo, Q. Wang, H. Xue, and B. Zhang, Higher-order Dirac semimetal in a photonic crystal, Phys. Rev. B **105**, L060101 (2022).

[21] J. Ningyuan, C. Owens, A. Sommer, D. Schuster, and J. Simon, Time- and Site-Resolved Dynamics in a Topological Circuit, Phys. Rev. X **5**, 021031 (2015).

[22] V. V. Albert, L. I. Glazman, and L. Jiang, Topological Properties of Linear Circuit Lattices,

Phys. Rev. Lett. **114**, 173902 (2015).

[23] S. Imhof *et al.*, Topolectrical-circuit realization of topological corner modes, Nat. Phys. **14**, 925 (2018).

[24] C. H. Lee, S. Imhof, C. Berger, F. Bayer, J. Brehm, L. W. Molenkamp, T. Kiessling, and R. Thomale, Topolectrical Circuits, Commun. Phys. 1, 39 (2018).

[25] M. Ezawa, Electric-circuit simulation of the Schrödinger equation and non-Hermitian quantum walks, Physical Review B 100, 165419 (2019).

[26] X.-X. Zhang and M. Franz, Non-Hermitian Exceptional Landau Quantization in Electric Circuits, Phys. Rev. Lett. 124, 046401 (2020).

[27] T. Helbig et al., Generalized bulk–boundary correspondence in non-Hermitian topolectrical circuits, Nat. Phys. 16, 747 (2020).

[28] M. Ezawa, Non-Abelian braiding of Majorana-like edge states and topological quantum computations in electric circuits, Phys. Rev. B 102, 075424 (2020).

[29] S. Liu, R. Shao, S. Ma et al., Non-Hermitian skin effect in a non-Hermitian electrical circuit, Research 2021, 5608038 (2021).

[30] W. Zhang, D. Zou, Q. Pei, W. He, H. Sun, and X. D. Zhang, Moiré circuits: Engineering magic-angle behavior, Phys. Rev. B 104, L201408 (2021).

[31] W. Zhang, H. Yuan, W. He, X. Zheng, N. Sun, F. Di, H. Sun and X.D. Zhang, Observation of interaction-induced phenomena of relativistic quantum mechanics, Commun. Phys. 4, 250 (2021).

[32] N. Pan, T. Chen, H. Sun, and X. D. Zhang, Electric-Circuit Realization of Fast Quantum Search, Research, Research 2021, 9793071 (2021).

[33] H. Zhang, Tian Chen, N. Pan, and X. D. Zhang, Electric-Circuit Simulation of Quantum Fast Hitting with Exponential Speedup, Adv. Quantum Technol. 2100143 (2022).

[34] J. Bao, D. Zou, W. Zhang, W. He, H. Sun, and X. D. Zhang, Topoelectrical circuit octupole insulator with topologically protected corner states, Phys. Rev. B **100**, 201406 (2019).

[35] T. Hofmann, T. Helbig, C. H. Lee, M. Greiter, and R. Thomale, Chiral Voltage Propagation and Calibration in a Topolectrical Chern Circuit, Phys. Rev. Lett. **122**, 247702 (2019).

[36] M. Ezawa, Electric circuit simulations of $n$th-Chern-number insulators in $2n$-dimensional space and their non-Hermitian generalizations for arbitrary $n$, Phys. Rev. B **100**, 075423 (2019).

[37] W. Zhang, D. Zou, J. Bao, W. He, Q. Pei, H. Sun, and X. D. Zhang, Topolectrical-circuit realization of a four-dimensional hexadecapole insulator, Phys. Rev. B **102**, 100102 (2020).

[38] R. Yu, Y. X. Zhao, and A. P. Schnyder, 4D spinless topological insulator in a periodic electric circuit, Natl. Sci. Rev. **7**, 1288 (2020).

[39] Y. Wang, H. M. Price, B. Zhang, and Y. D. Chong, Circuit implementation of a four-dimensional topological insulator, Nat. Commun. **11**, 2356 (2020).

[40] N. A. Olekhno *et al.*, Topological edge states of interacting photon pairs emulated in a topolectrical circuit, Nat. Commun. **11**, 1436 (2020).

[41] C. H. Lee *et al.*, Imaging nodal knots in momentum space through topolectrical circuits, Nat. Commun. **11**, 4385 (2020).

[42] R. Li, B. Lv, H. Tao, J. Shi, Y. Chong, B. Zhang, and H. Chen, Ideal type-II Weyl points in topological circuits, Natl. Sci. Rev. **8** (2020).

[43] W. Zhang, D. Zou, Q. Pei, W. He, J. Bao, H. Sun, and X. D. Zhang, Experimental Observation of Higher-Order Topological Anderson Insulators, Phys. Rev. Lett. **126**, 146802 (2021).

[44] D. Zou, T. Chen, W. He, J. Bao, C. H. Lee, H. Sun, and X.D. Zhang, Observation of hybrid higher-order skin-topological effect in non-Hermitian topolectrical circuits, Nat. Commun. **12**, 7201 (2021).

[45] T. Helbig, T. Hofmann, C. H. Lee, R. Thomale, S. Imhof, L. W. Molenkamp, and T. Kiessling, Band structure engineering and reconstruction in electric circuit networks, Phys. Rev. B **99** (2019).